\documentclass{article}
\usepackage{amsmath}
\usepackage{CJK}
\usepackage{indentfirst}  
\protect
\begin{document}

\begin{CJK}{GBK}{fs}

\title{
      Symmetric Space-times from Thermodynamics}
\author{\small{Jinbo Yang, Hongwei Tan, Tangmei He, Jingyi Zhang\footnote{{Corresponding author, E-mail:
physicz@aliyun.com}} }\\ \small{\textit{Centor for Astrophysics,Guangzhou University,510006,Guangzhou,China}}}

\maketitle
\begin{abstract}
  We apply thermodynamics method to generate exact solution with maximum symmetric surface for Einstein equation without solving it. 
  The exact solutions are identified with which people have solved before. 
  The horizons structure of solutions are discussed in different situation. 
  These results suggest that plane,spherical,pseudo-spherical symmetric spacetime form a family of solutions, 
   and this relationship can also be extended to Kerr spacetime and Taub-NUT spacetime. 
  It sheds light on exploring more general definition of Misner-Sharp mass.\\
  Key words: Misner-Sharp mass,Thermodynamics, Maximum symmetric ,Exact solutions, Kerr family , Taub-NUT family
  \\\textit{PACS}: 04.70.Dy
\end{abstract}



\section{Introduction}
In 1970s, the similarity between four laws of black hole mechanics and four laws of  thermodynamics was discovered\cite{BlackHoleLaws,PhysRevD.7.2333}.
Later, the discovery of Hawking radiation made them more closely linked\cite{hw2,Hrad1}.
And it inspired people to explore deeper connection between gravity and thermodynamics.
More concretely, people found that Einstein equation could be derived through thermodynamics consideration\cite{PhysRevLett.75.1260,gravfromthermo}.
Further,the idea about entropy force was proposed and got development\cite{entroVerlin,PhysRevD.81.084012}.
Their ideas might partially base on Killing vector field, which meant that they consider the static situation. But a unified first law was discover in un-static spacetime base on the concept of Misner-Sharp mass(MS mass)\cite{UfLbyHeyward}. And it had application in cosmology \cite{PhysRevD.75.064008}.

Recently,people found that the Schwarzchild solution, RN solution, Schwarzchild-(Anti)de Sitter solution and their higher dimension generalization could be derived from thermodynamics directly rather than solving Einstein equation \cite{PhysRevD.89.064052}.
The derivation is also base on the concept of Misner-Sharp mass(MS mass).
The definition of MS mass is just for spherical symmetric spacetime in general theory of relativity at the beginning\cite{PhysRev.136.B571}.
Latter ,the definition was generalized to other gravity theory in 2 codimension maximum symmetric spacetime,just like Gauss-Bounet gravity\cite{PhysRevD.77.064031},4 dimension f(R) theory\cite{PhysRevD.80.104016}, n dimension f(R)\cite{PhysRevD.90.024062}, etc.
Then some exact solution were derived.
For example,topological Schwarzchild solution in 4 dimension GR, include spherical symmetric, plane-symmetric and pseudo-spherical symmetric.
And exact solutions for high dimension $f(R)$ theory in the case of $f(R)=R^d$ can also be gain in this way\cite{PLB737}.
Actually, the same method can also be applied to generate solutions with $U(1)$ charge and cosmological constant $\Lambda$.
Some solution which have been derived by solving Einstein equation would be gain simply by this method.
Inversely speaking,thermodynamics method is so powerful such that the corresponding spherical symmetric, plane-symmetric and pseudo-spherical symmetric spacetime can form a family of exact solutions.
It shows the deep connection between gravity and thermodynamics.

This paper will be organized in this way.
Section 2 will introduce how to get 2 codimension maximum symmetric solution with $U(1)$ charge and cosmological constant $\Lambda$ through the thermodynamics method.
Section 3 will discuss the plane-symmetric case.
We identify the solution with plane-symmetric which have been got by solving Einstein equation before.
Section 4 will discuss the pseudo-spherical symmetric case.
Finally,we will make a discussion in section 5.

\section{Exact Solution from Thermodynamics}
 Considering static spacetime with 2 dimension maximum symmetric surface, no loss general,we give this metric ansatz.
 \begin{equation}
   ds^2=-f(r)dt^2+\frac{dr^2}{h(r)}+r^2d\Omega_2^2
 \end{equation}
 Generalized MS mass for spacetime with maximum symmetric 2-surface is defined as \cite{PLB737}.
 \begin{equation}
   M_{ms}(r)=\frac{r}{2}(k-h(r))
 \end{equation}
 In spherical symmetric case, $d\Omega_2^2=d\theta^2+\sin^2\theta d\varphi^2$,$k=1$.
 In plane-symmetric case $d\Omega_2^2=dx^2+dy^2$, $k=0$.
 In pseudo-spherical symmetric case $d\Omega_2^2=d\theta^2+\sinh^2\theta d\varphi^2$, $k=-1$.
 Here we would apply the thermodynamics method to generate solution of Einstein equation. Make a variation with respective to r
  \[   \delta M_{ms}(r)=\frac{k-h(r)-rh'(r)}{2}dr \]
 Consider the first law in an adiabatic Misner-Sharp system with $\Lambda$ term and $q$ term
 \footnote{
 The original term about charge $\frac{q}{r}dq$ can't derive RN solution.
 It should be corrected as $\frac{q^2}{2r^2}$.}.
 \begin{equation}
   \delta M_{ms}(r)=\frac{\Lambda}{2} r^2dr+\frac{q^2}{2r^2}dr
 \end{equation}
  Then,we have equation
  \begin{equation}
   k-h(r)-rh'(r)=\Lambda r^2+\frac{q^2}{r^2}
 \end{equation}
  The solution is
  \begin{equation}
    h(r)=k-\frac{C}{r}+\frac{q^2}{r^2}-\frac{\Lambda r^2}{3}
 \end{equation}
 $C$ is the integration constant. Substitute $h(r)$ into the formula of MS mass,we get
 \begin{equation}
     M_{ms}(r)=\frac{r}{2}(k-(k-\frac{C}{r}+\frac{q^2}{r^2}-\frac{\Lambda r^2}{3}))=\frac{C}{2}-\frac{q^2}{2r}+\frac{\Lambda r^3}{6}
 \end{equation}

 On the one hand ,$h(r)$ can be solved by above method. On the other hand,the relationship between Killing surface gravity and geometry surface gravity should be applied here in order to find $f(r)$.Under the metric ansatz,Killing surface gravity is
  \begin{equation}
     \kappa(r)=\frac{1}{2}\sqrt{\frac{h}{f}}f'
 \end{equation}
 Base on unified first law \cite{UfLbyHeyward,PhysRevD.89.064052}, define geometry surface gravity ,$w$ is the work term.Here is the definition.
  \begin{equation}
     w=-\frac{1}{2}I^{ab}T_{ab}
 \end{equation}
 Demand Killing surface gravity equal to geometry surface gravity,then
 \begin{equation}
     \kappa(r)=\frac{M_{ms}}{r^2}-4\pi rw
 \end{equation}
 Work terms with cosmological constant and charge are given here \cite{PhysRevD.89.064052}
  \begin{equation}
     w_{\Lambda}=\frac{\Lambda}{8\pi}\;,\; w_{q}=\frac{q^2}{8\pi r^4}
 \end{equation}
 Then, the solution is
 \begin{equation}
   f(r)=(\sqrt{k-\frac{C}{r}+\frac{q^2}{r^2}-\frac{\Lambda r^2}{3}}+D )^2
 \end{equation}
 Set $D=0$,we have
 \begin{equation}
   ds^2=-(k-\frac{C}{r}+\frac{q^2}{r^2}-\frac{\Lambda r^2}{3})dt^2+\frac{dr^2}{k-\frac{C}{r}+\frac{q^2}{r^2}-\frac{\Lambda r^2}{3}}+r^2d\Omega_2^2
 \end{equation}
Generally, the vector-potential of electric-magnetic field is $A=\frac{q}{r}dt$.
Putting it into Einstein equation and Maxwell equation for checking,
it is indeed electromagnetic vacuum solution with cosmological constant.
 If $k=1$,it's just RN- (A)dS black hole.
 And we can see that not only spherical symmetric solutions, but also plane-symmetric solution and pseudo-spherical symmetric solutions with charge and cosmological constant can be derived in this simple way.
 \footnote{ Set $C=q=\Lambda=0$ except $k=0$ case, these solutions would go back to  Minkowski spacetime. }

 It's worth noting that the sign of functions $f(r)$ and $h(r)$ can be negative.
 In this method, there are hidden assumptions $f(r)>0$,$h(r)>0$.
 However,conditions $f(r)<0$ and $h(r)<0$ are also satisfied Lorentz signature.
 It can be checked that it's still the solution of Einstein's theory.
 Further more, generally we have $f(r)=h(r)$, so the requirement of Lorentz signature is satisfied automatically.
 Thus, in later parts of our paper, we will extend our result to situations about $f(r)<0$, $h(r)<0$.
\section{Plane-symmetric Solutions}
  In this section,  we would discuss $k=0$ solutions in detail.
  These are Plane-symmetric cases.
  They are corresponding to solutions which are already got by solving Einstein equation,just like Taub spacetime, $AdS_4$ black hole, etc.
  Let's consider the solution without $q$ or $\Lambda$ first.
  The metric is
\begin{equation}
   ds^2=-(-\frac{C}{r})dt^2
         +\frac{dr^2}{-\frac{C}{r}}+r^2(dx^2+dy^2)
 \end{equation}
 $C$ cannot be 0, then it is Taub solution\cite{TaubPlane}.
 If $C<0$, redefine the parameter $C=-2\mu$. Here $\mu>0$, it's a constant with mass dimension.Then
 \begin{equation}
   ds^2=-\frac{2\mu}{r}dt^2
         +\frac{dr^2}{\frac{2\mu}{r}}+r^2(dx^2+dy^2)  \label{VarTaub}
 \end{equation}
 Introduce coordinate transformation $z=\frac{r^2}{4\mu}$. Thus, the metric in new coordinate frame is
 \begin{equation}
   ds^2=\sqrt{\frac{\mu}{z}}(-dt^2+dz^2)+4\mu z(dx^2+dy^2)
 \end{equation}
   In fact, it's static Taub spacetime. Make scale transformation
   $ t\rightarrow \mu^{-\frac{1}{3}}t $,$ z\rightarrow \mu^{-\frac{1}{3}}z $
   $ x\rightarrow\frac{1}{2} \mu^{-\frac{1}{3}}x $,
   $ y\rightarrow\frac{1}{2} \mu^{-\frac{1}{3}}y $ ,
   a more familiar line-element is found as
 \begin{equation}
   ds^2=\sqrt{\frac{1}{z}}(-dt^2+dz^2)+z(dx^2+dy^2)
 \end{equation}
 If $C>0$, as a mater of fact, it is spatial homogenous but un-static Taub solution.
 Similarly, the coordinate frame which make metric have following form can be found.
 \begin{equation}
   ds^2=\sqrt{\frac{1}{z}}(dt^2-dz^2)+z(dx^2+dy^2)
 \end{equation}
 Then, two situation of Taub spacetime are corresponding to  $C>0$ and $C<0$.

 Plane-symmetric solution without cosmological constant but with charge is \cite{OLDKar,PhysRevD.27.1731}
 \begin{equation}
   ds^2=-(-\frac{C}{r}+\frac{q^2}{r^2})dt^2+\frac{dr^2}{-\frac{C}{r}
          +\frac{q^2}{r^2}}+r^2(dx^2+dy^2)
 \end{equation}
  Define $ dz=\frac{dr}{-\frac{C}{r}+\frac{q^2}{r^2}}$ ,so
  \begin{equation}
   -2Cz=(r+\frac{q^2}{C})^2+2(\frac{q^2}{C})^2\log|r-\frac{q^2}{C}|
 \end{equation}
 It's the solution found by Letelier and Tabensky in 1974 \cite{LTplane}. Their work is based on the foundation of Patnaik \cite{PatnakPlane}.

 With the same method, we can define $C=-2\mu$ if $C<0$. When $r>0$,$\frac{2\mu}{r}+\frac{q^2}{r^2}>0$,
 so there is no horizon in this case.
 If $C>0$, refind $C=2M$,then in the range of $0<r<\frac{q^2}{2M}$, there is $-\frac{2M}{r}+\frac{q^2}{r^2}>0$. So $r=\frac{q^2}{2M}$ is the Killing horizon.

  If charge $q=0$ but $\Lambda\neq0$, there are more complex structure.
 \begin{equation}
   ds^2=-(-\frac{C}{r}-\frac{\Lambda r^2}{3})dt^2
         +\frac{dr^2}{-\frac{C}{r}-\frac{\Lambda r^2}{3}}+r^2(dx^2+dy^2)
 \end{equation}
 If $C>0,\Lambda>0$, it's a spatial homogenous spcaetime. If $C<0,\Lambda>0$ ,demand $C=-\frac{r_h^3}{L^2},\frac{\Lambda}{3}=\frac{1}{L^2}$, $r_h>0$ then
 \begin{equation}
   h(r)=\frac{r_h^3}{rL^2}-\frac{r^2}{L^2}
 \end{equation}
 Obviously, $r=r_h$ is Killing horizon. In the range of $0<r<r_h$,this metric describe a static part of the spacetime.
 Region behind the horizon is spatial homogenous.

 Similarly,if $C<0,\Lambda<0$, it is a static spacetime without Killing horizon.If $C>0,\Lambda<0$,there is a horizon. Set $C=\frac{r_h^3}{L^2},\frac{\Lambda}{3}=-\frac{1}{L^2}$,$r_h>0$,thus
 \begin{equation}
    \frac{r^2}{L^2}-\frac{r_h^3}{rL^2}
 \end{equation}
 Obviously,$r=r_h$ is Killing horizon. In the range of $r>r_h$, it's a static region.
 Actually, this is so called $AdS_4$ Schwarzchild black hole with planar horizon which is applied in holographic superconductor \cite{PhysRevLett.101.031601}.

 The most complicated case is that all parameters are not equal to 0.
 It is also applied in the research of holographic superconductor\cite{JHEPcai2} and also be the general case of plane-symmetric solution given by Ref.\cite{PhysRevD.27.1731}.
 \begin{equation}
   ds^2=-(-\frac{C}{r}+\frac{q^2}{r^2}-\frac{\Lambda r^2}{3})dt^2+\frac{dr^2}{-\frac{C}{r}+\frac{q^2}{r^2}-\frac{\Lambda r^2}{3}}+r^2(dx^2+dy^2)
 \end{equation}

  If $\Lambda<0$,define $-\frac{\Lambda}{3}=\frac{1}{L^2}$. Then,make a scale transformation $r\rightarrow rL$ and redefine all parameter, $C\rightarrow CL$, $q\rightarrow qL$.
The equation $f(r)=0$ leads to
 \begin{equation}
   \frac{-Cr+q^2+ r^4}{r^2}=0
 \end{equation}
 It's impossible to have 4 positive root due to the lack of $r^3$ term.
 Since just two parameter $C$ and $q$, assume $r=a$ and $r=b$ are real roots of the equation, then $C$ and $q$ can be expressed by them.
 Write down the formula
 \begin{equation}
   r^4-Cr+q^2=(r-a)(r-b)(r^2+(a+b)r+a^2+ab+b^2)
 \end{equation}
 Compare the term with same order,then
  \begin{equation*}
   C=a^3+a^2b+ab^2+b^3
 \end{equation*}
 \begin{equation}
   q^2=a^3b+a^2b^2+ab^3
 \end{equation}
 Further more, assume $r=a$ and $r=b$ are positive.
 Under this condition, whether other real roots exist depend on equation
 $r^2+(a+b)r+a^2+ab+b^2=0$.
 Obviously, $(a+b)^2-4 (a^2+ab+b^2)=-3 a^2 - 2 a b - 3 b^2<0 $.
 So it's even impossible to have real root.
 We can claim that there are no more then two Killing horizons in this case.
 No loss general, assume $a>b$, then the region $r>a$ and $0<r<b$ are static, while the region $b<r<a$ is spatial homogenous un-static.

 Through similar method, we found that there is no more then one Killing horizon  when $\Lambda>0$.
 Suppose the horizon is located at $r=r_h$,
 In the range of $r>r_h$, this region is un-static but spatial homogenous,
 while $r<r_h$ is static region.
 \section{Pseudo-spherical Symmetric Spacetime}
 If the 2-dimension maximum symmetric surface is pseudo-sphere,then we have this solution
 \begin{equation}
   ds^2=-(-1-\frac{C}{r}+\frac{q^2}{r^2}-\frac{\Lambda r^2}{3})dt^2+\frac{dr^2}{-1-\frac{C}{r}+\frac{q^2}{r^2}-\frac{\Lambda r^2}{3}}+r^2(d\theta^2+\sinh^2\theta d\varphi^2)
 \end{equation}
 The spacetime described by this metric is the general situation with $k=-1$.
 And the case of $q=\Lambda=0$ is called topological Schwarzchild spacetime in Ref.\cite{PLB737}, which has already been got in Ref.\cite{PSSSch}.
 Pseudo-spherical symmetric solution with negative cosmological constant is presented in literatures \cite{PhysRevD.82.044034,PhysRevD.91.124039}.

 The simplest case
 is
 \begin{equation}
   ds^2=-(-1-\frac{C}{r})dt^2+\frac{dr^2}{-1-\frac{C}{r}}+r^2(d\theta^2+\sinh^2\theta d\varphi^2)
 \end{equation}
 If $C<0$, there is one horizon, region inside the horizon is static. Instead, outside the horizon is un-static.
 It has been given in \cite{PSSSch,PseudoWormhole}.In \cite{PseudoWormhole},  the $k=-1$ tranvalable wormhole was found.

 Solution with charge but without cosmological constant is
 \begin{equation}
   ds^2=-(-1-\frac{C}{r}+\frac{q^2}{r^2})dt^2+\frac{dr^2}{-1-\frac{C}{r}+\frac{q^2}{r^2}}
   +r^2(d\theta^2+\sinh^2\theta d\varphi^2)
 \end{equation}
 Then we just discuss its horizons structure,
 roots of equation $-1-\frac{C}{r}+\frac{q^2}{r^2}=0$ are
 \begin{equation}
   r_{+}=-\frac{C}{2}+\sqrt{(\frac{C}{2})^2+q^2} \;,\;
   r_{-}=-\frac{C}{2}-\sqrt{(\frac{C}{2})^2+q^2}
 \end{equation}
 No matter the sign of $C$, it's just one positive root $r_{+}$.
 Thus, the region $0<r<r_{+}$ is static, while $r>r_{+}$ is un-static but spatial homogenous.

General solution with charge and cosmological constant may be a new solution.
  \begin{equation}
   ds^2=-(-1-\frac{C}{r}+\frac{q^2}{r^2}-\frac{\Lambda r^2}{3})dt^2+\frac{dr^2}{-1-\frac{C}{r}+\frac{q^2}{r^2}-\frac{\Lambda r^2}{3}}+r^2(d\theta^2+\sinh^2\theta d\varphi^2)
 \end{equation}
 When $\Lambda<0$, things are quite different.
 Similar to what we do for plane-symmetric solution,assume
 $r=a$ and $r=b$ are real roots, thus
 \begin{equation}
   r^4-r^2-Cr+q^2=(r-a)(r-b)(r^2+(a+b)r+a^2+ab+b^2-1)
 \end{equation}
 And
 \[C= a^3 + a^2 b+a b^2+ b^3 -(a+b)  \]
 \begin{equation}
   q^2=a^3 b + a^2 b^2 + a b^3 - a b
 \end{equation}
 $a$ and $b$ cannot both be positive unless condition $a^2+b^2+ab>1$ is satisfied. However, this condition also lead to that equation $r^2+(a+b)r+a^2+ab+b^2-1=0$ has no positive roots.
 Then we can claim that there are no more than two horizons in this case. The results is similar to plane-symmetric solutions.
 Assume $a>b$,
 then region $r>a$ and $0<r<b$ are static,
 while region $b<r<a$ is spatial homogenous un-static.
 But one interesting thing is that when $q=0$, it may still have two positive roots. Still assume they are $a$ and $b$, then $a^2+b^2+ab$ should equal to 1.

  If $\Lambda>0$, we can define $\frac{\Lambda}{3}=\frac{1}{L^2}$,
  and make the same scale transformation. Now, there are
  \[r^4+r^2+Cr-q^2=(r-a)(r-b)(r^2+(a+b)r+a^2+ab+b^2-k)\]
 \[C=- a^3 - a^2 b- a b^2- b^3-a-b \]
\begin{equation}
  q^2=-a^3 b - a^2 b^2 -a b^3 - a b
 \end{equation}
 There is no more then one positive root due to the positivity of $q^2$. The results is also similar to plane-symmetric solutions.
 It is un-static outside,
 but static inside.
\section{Discussion}

 In this paper,we apply MS mass and unified first law to derive a family of 2 codimension maximum symmetric solution in electromagnetic vacuum with cosmological constant,include famous solution like Schwarzchild-(anti) de Sitter solution, RN-(anti) de Sitter solution, Taub solution, $AdS_4$ black hole,etc.
 There are 3 subclasses according to $k$.
 Concretely, $k=1$ correspond to spherical symmetry, $k=0$ correspond to plane-symmetry, $k=-1$ correspond to pseudo-spherical symmetry.
 In every class, $C,q$ and $\Lambda$ determine the horizon structure.
 It's worth noting that all exact solutions for $k=1$, $k=0$ and $k=-1$ form a family whith pattern like this:
 Change $\sin\theta $ to $\sinh\theta$ while $k=1$ to $k=-1$, pseudo-spherical symmetric solution can be generated from spherical symmetric solution.
 For plane symmetric solution, change $\sin\theta $ to $\theta$ while $k=1$ to $k=0$, and interpret $\theta$ as radius coordinate on the plane.
 Take $x=\theta\cos\varphi,y=\theta\sin\varphi$, we have $d\theta^2+\theta^2 d\varphi^2=dx^2+dy^2$.
 We are inspired by this pattern.
    Although the thermodynamics method cannot derive Kerr solution, however, it's well known that Kerr spacetime can be regarded as the solution deformed from the simplest spherical symmetric solution called Schwarchild spacetime.
    We have learned that there are plane symmetric solution called Taub spacetime and  pseudo-spherical symmetric solution called topological Schwarzchild spacetime or anti-Schwarzchild spacetime.
    In principle, they can also generate corresponding spinning solution like Kerr spacetime.
    We do find them out here. Firstly, Kerr spacetime is given here
    \begin{equation}
     \begin{split}
        ds^2=&-(1-\frac{2Mr}{\rho^2})dt^2
           +\frac{\rho^2}{r^2+a^2-2Mr}dr^2
           +\rho^2d\theta^2 \\
           &+(r^2+a^2+\frac{2Mr}{\rho^2}a^2\sin^2\theta)\sin^2\theta d\varphi^2
           -2\frac{2Mr}{\rho^2}a \sin^2\theta dt d\varphi
        \end{split}
   \end{equation}
   In this formula , $\rho^2=r^2+a^2\cos^2\theta $.
   Then, in order to extend to plane symmetric case, change some details by following rules.
   Change $1$ in front of $dt^2$ and $r^2+a^2$ under $dr^2$ to $0$; change $\cos\theta $ to $1$ in $\rho^2$ and replace every $\sin\theta $ by $\theta$, we have
   \begin{equation}
     \begin{split}
        ds^2=&\frac{2Mr}{\rho^2}dt^2
           -\frac{\rho^2}{2Mr}dr^2
           +\rho^2d\theta^2 \\
           &+(r^2+a^2+\frac{2Mr}{\rho^2}a^2\theta^2)\theta^2 d\varphi^2
           -2\frac{2Mr}{\rho^2}a\theta^2 dt d\varphi     \label{planeKerr}
        \end{split}
   \end{equation}
   Calculating the Ricci tensor and $R^{\lambda\sigma\mu\nu}R_{\lambda\sigma\mu\nu}$, the result shows that $R_{\mu\nu}=0$ and $R^{\lambda\sigma\mu\nu}R_{\lambda\sigma\mu\nu} \propto \rho^{-12}$.
   It shows that this is indeed the exact solution of vacuum Einstein equation.
   This modify-by-hand method seems questionable, but one can find the generalized Newman complex transformation from Exp.(\ref{VarTaub}).
   First, use polar coordinate and replace $\mu$ by $-M$, then introduce the Eddington coordinate $du=dt-dr/(\frac{-2M}{r})$, to rewrite the line element (\ref{VarTaub}).
   \begin{equation}
     ds^2=\frac{2M}{r}du^2-2dudr+r^2d\theta^2+r^2\theta^2d\phi^2
   \end{equation}
   Select tetrad as follow
   \begin{equation}
    \begin{split}
      &k^a=\frac{1}{\sqrt2}(\frac{\partial}{\partial r})^a   \\
      &l^a=\frac{1}{\sqrt2}((\frac{\partial}{\partial
       u})^a-\frac{f}{2}(\frac{\partial}{\partial r})^a)   \\
      &m^a=\frac{1}{\sqrt2r}((\frac{\partial}{\partial \theta})^a
                 -\frac{i}{\theta}(\frac{\partial}{\partial\varphi})^a)\\
      &\bar{m}^a=\frac{1}{\sqrt2r}((\frac{\partial}{\partial \theta})^a
                 +\frac{i}{\theta}(\frac{\partial}{\partial\varphi})^a)
    \end{split}
   \end{equation}
   Then, the generalized Newman complex transformation for plane symmetric is
   \begin{equation}
     u'=u-ia\frac{\theta^2}{2} \;,\; r'=r-ia
   \end{equation}
  So, $du'=du-ia\theta d\theta$ and $dr'=dr$. Substitute them into the line element,

   and introduce $du'=dt-\frac{2Mr}{r^2+a^2}dr$, then one can get Exp.(\ref{planeKerr})

   And there is no singularity for this spacetime since $R^{\lambda\sigma\mu\nu}R_{\lambda\sigma\mu\nu}$ is finite everywhere due to $\rho^{2}\geq a^2$.
   For pseudo symmetrical solution, we apply similar rules and get
   \begin{equation}
     \begin{split}
        ds^2=&-(-1-\frac{2Mr}{\rho^2})dt^2
           +\frac{\rho^2}{-(r^2+a^2)-2Mr}dr^2
           +\rho^2d\theta^2 \\
           &+(r^2+a^2+\frac{2Mr}{\rho^2}a^2sh^2\theta)sh^2\theta d\varphi^2
           -2\frac{2Mr}{\rho^2}a sh^2\theta dt d\varphi
        \end{split}  \label{hyperKerr}
   \end{equation}
   In this formula,$\rho^2=r^2+a^2ch^2\theta$.
   Similarly, there is $R_{\mu\nu}=0$ and $R^{\lambda\sigma\mu\nu}R_{\lambda\sigma\mu\nu} \propto \rho^{-12}$.
   It is also singularity-free since $\rho^{2}\geq a^2$.
   This is so called S-Kerr solution\cite{RegularSBrane04,NonSingSbrane04} which is one kind of S-brane\cite{S-Branes2002}.
   The the generalized Newman complex transformation to derive it is as following.
   Let's start with pseudo spherical symmetric Schilwarzchild solution under the Eddington coordinate $du=dt-dr/(-1-\frac{2M}{r})$,
   \begin{equation}
     ds^2=-(-1-\frac{2M}{r})du^2-2dudr+r^2d\theta^2+r^2\sinh\theta^2d\phi^2
   \end{equation}
   The tetrad is took as
   \begin{equation}
    \begin{split}
      &k^a=\frac{1}{\sqrt2}(\frac{\partial}{\partial r})^a   \\
      &l^a=\frac{1}{\sqrt2}((\frac{\partial}{\partial
       u})^a-\frac{f}{2}(\frac{\partial}{\partial r})^a)   \\
      &m^a=\frac{1}{\sqrt2r}((\frac{\partial}{\partial \theta})^a
                 -\frac{i}{\sinh\theta}(\frac{\partial}{\partial\varphi})^a)\\
      &\bar{m}^a=\frac{1}{\sqrt2r}((\frac{\partial}{\partial \theta})^a
                 +\frac{i}{\sinh\theta}(\frac{\partial}{\partial\varphi})^a)
    \end{split}
   \end{equation}
   Then, the generalized Newman complex transformation for plane symmetric is
   \begin{equation}
     u'=u-ia\cosh\theta \;,\; r'=r-ia\cosh\theta
   \end{equation}
   So, $du'=du-ia\theta d\theta$ and $dr'=dr-ia\sinh\theta$. Substitute them into the line element, and introduce $du'=dt-\frac{2Mr}{r^2+a^2}dr$, then one can get Exp.(\ref{hyperKerr}).
   These results show that there is also a Kerr family with members $k=1$, $k=0$ and $k=-1$.
   The relationship  can also be extended to spacetime with charge or cosmological constant.
   The similarity between $k=1$ , $k=0$ and $k=-1$ solutions can be also extend to Taub-NUT solution. Taub-NUT solution is
   \begin{equation}
    \begin{split}
        ds^2=&U^{-1}(r)dr^2
           -U(r)(dt+2l\cos\theta d\phi)^2 \\
            &+(r^2+l^2)(d\theta^2+\sin^2\theta d\phi)^2
        \end{split}
   \end{equation}
   in which  $U(r)=1-2\frac{m r +l^2}{r^2+l^2}$.
   There is also a generalized Newman complex transformation to generate it from Schwarzchild solution.
   Take the tetrad
    \begin{equation}
    \begin{split}
      &k^a=\frac{1}{\sqrt2}((\frac{\partial}{\partial
       t})^a+U(\frac{\partial}{\partial r})^a)   \\
      &l^a=\frac{1}{\sqrt2}(\frac{1}{U}(\frac{\partial}{\partial
       t})^a-(\frac{\partial}{\partial r})^a)   \\
      &m^a=\frac{1}{\sqrt2r}((\frac{\partial}{\partial\theta})^a
                 -\frac{i}{\sin\theta}(\frac{\partial}{\partial\varphi})^a)\\
      &\bar{m}^a=\frac{1}{\sqrt2r}((\frac{\partial}{\partial\theta})^a
                 +\frac{i}{\sin\theta}(\frac{\partial}{\partial\varphi})^a)
   \end{split}
   \end{equation}
  Then introduce
  \begin{equation}
  t'=t+2il\log{(\sin\theta)} \;,\;  r'=r+il
  \end{equation}
  Such that the Taub-Nut solution can be derived.

   For pseudo-symmetric case, the tetrad is
   \begin{equation}
    \begin{split}
      &k^a=\frac{1}{\sqrt2}((\frac{\partial}{\partial
       t})^a+U(\frac{\partial}{\partial r})^a)   \\
      &l^a=\frac{1}{\sqrt2}(\frac{1}{U}(\frac{\partial}{\partial
       t})^a-(\frac{\partial}{\partial r})^a)   \\
      &m^a=\frac{1}{\sqrt2r}((\frac{\partial}{\partial\theta})^a
                 -\frac{i}{\sinh\theta}(\frac{\partial}{\partial\varphi})^a)\\
      &\bar{m}^a=\frac{1}{\sqrt2r}((\frac{\partial}{\partial\theta})^a
                 +\frac{i}{\sinh\theta}(\frac{\partial}{\partial\varphi})^a)
   \end{split}
   \end{equation}
   The complex transformation is
   \begin{equation}
     t'=t+2il\log{(\sinh\theta)} \;,\;  r'=r+il
   \end{equation}
   Then, the following metric can be derived
   \begin{equation}
    \begin{split}
        ds^2=&U^{-1}(r)dr^2
           -U(r)(dt+2l\cosh\theta d\phi)^2 \\
            &+(r^2+l^2)(d\theta^2+\sinh^2\theta d\phi)^2
        \end{split}
   \end{equation}
   in which $U(r)=-1+2\frac{m r +l^2}{r^2+l^2}$.

  Finally, for plane symmetric case, the tetrad is
  \begin{equation}
    \begin{split}
      &k^a=\frac{1}{\sqrt2}((\frac{\partial}{\partial
       t})^a+U(\frac{\partial}{\partial r})^a)   \\
      &l^a=\frac{1}{\sqrt2}(\frac{1}{U}(\frac{\partial}{\partial
       t})^a-(\frac{\partial}{\partial r})^a)   \\
      &m^a=\frac{1}{\sqrt2r}((\frac{\partial}{\partial\theta})^a
                 -\frac{i}{\theta}(\frac{\partial}{\partial\varphi})^a)\\
      &\bar{m}^a=\frac{1}{\sqrt2r}((\frac{\partial}{\partial\theta})^a
                 +\frac{i}{\theta}(\frac{\partial}{\partial\varphi})^a)
   \end{split}
   \end{equation}
  The complex transformation is
  \begin{equation}
     t'=t+il\frac{\theta^2}{2} \;,\; r'=r+il
   \end{equation}
  It's very interesting that plane symmetric solution in Taub-NUT family is again the $k=0$ solution Exp.(\ref{planeKerr}) in Kerr family.

  These results show that thermodynamics method ties different solutions of Einstein equation together in a simple way and also give us some hints to generalized the concept of Misner-Sharp mass for non-maximum symmetric spcetime like Kerr spacetime and Taub-NUT spacetime.
\section*{Acknowledgments}
This research is supported by the National Natural Science Foundation of China under Grant Nos 11273009 and 11303306.

\end{CJK}
\end{document}